\documentclass{aa}

\def\degr{$^{\circ}$}

\usepackage{graphicx}
\usepackage{natbib}
\bibpunct{(}{)}{;}{a}{}{,}

\usepackage{times}
\begin{document}
\title{Screening the Hipparcos-based astrometric orbits of sub-stellar 
objects\thanks{Based on observations from the Hipparcos astrometric 
satellite 
operated by the European Space Agency (ESA 1997)}}
\titlerunning{Hipparcos and sub-stellar objects II}
\author{D.~Pourbaix\thanks{Postdoctoral Researcher, F.N.R.S., 
Belgium}\inst{1,2} 
\and F.~Arenou\inst{3}}
\institute{Institut d'Astronomie et d'Astrophysique, Universit\'e Libre de 
Bruxelles, C.P.~226, Boulevard du Triomphe, B-1050 Bruxelles
\and
Department of Astrophysical Sciences, Princeton University, 
Princeton NJ 08544-1001, U.S.A.
\and
UMR 8633 CNRS/DASGAL, Observatoire de Paris, 5 place J.~Janssen, 
F-92195 Meudon Cedex}
\date{Received date; accepted date} 
\offprints{D.~Pourbaix \email{pourbaix@astro.ulb.ac.be}}
\abstract{
The combination of Hipparcos astrometric
data with the spectroscopic data of putative extrasolar planets 
seems to indicate that a significant fraction of these low-mass
companions could be brown or M dwarfs \citep{Han-2001:a}.
We show that this is due to the adopted reduction procedure, and
consequently that the Hipparcos data do not reject the planetary 
mass hypothesis in all but one cases. Additional companions, undetected
so far, might also explain the large astrometric residuals of some of
these stars.
}
\maketitle

%
\section{Introduction}
%

The first attempt to combine the spectroscopic information
concerning extrasolar planetary candidates with the Hipparcos data
has been done by \cite{Perryman-1996:a}. The goal was to derive upper 
mass limits while the radial velocity analysis gives the lower limit.

\cite{Han-2001:a} lately investigated the mass of 30 extrasolar 
companions with period longer than 10 days using the Hipparcos 
\citep{Hipparcos} data.  For 27 systems, the inclination they derive 
is smaller than 20\degr\ (in 8 cases, it is even less than 1\degr), 
and they conclude that half of their sample stars might have brown or M 
dwarf secondary.

In this paper, we first derive similar inclinations for an extended set
of systems with no limitation based
on the period.  Such a large percentage of small inclinations is very 
unlikely (Sect.~\ref{Sect:incli}).  Instead of explaining it with an 
observational bias of radial velocity surveys \citep{Han-2001:a}, we 
show that the fitting procedure is responsible for the bias.  Another 
model is then used to fit the observations.  The comparison between 
the two sets of solutions allows us to conclude that the data seldom 
contain enough information to derive a reliable value of the 
inclination and of the semi-major axis.  Such a poor reliability will 
be illustrated in Sect.~\ref{Sect:HD209458}.

Although we show in Sect.~\ref{Sect:incl} that the derived 
inclinations are strongly doubtful, we expect a very small number of 
small inclinations in the putative extrasolar planets sample, and the 
problem is therefore to distinguish between a true astrometric motion and 
the effect of random noise.  Statistical tests are then proposed in 
Sect.~\ref{Sect:screen} in order to detect the real massive 
secondaries, not due to the effect of random noise or to the 
perturbation of other companions.

%
\section{Statistical properties of the low-mass companions sample}\label{Sect:stat}
%
%
\subsection{The inclinations}\label{Sect:incli}
%

The method adopted for fitting the astrometric data \citep{Mazeh-1999:a,
Halbwachs-2000:a,Han-2001:a,Zucker-2000:a} generally consists in fixing 
$K_1$ (reckoned in m/s) and the 
orbital parameters $\omega_1$, $e$, $P$ (reckoned in days), and $T$ 
to their spectroscopic values and to impose
\begin{equation}
a_a\sin i=9.19\,10^{-8}K_1 P\sqrt{1-e^2}\varpi\label{Eq:defaasini}
\end{equation}
The parallax, $\varpi$, reckoned in mas is adopted from Hipparcos
and $a_a$ is the semi-major axis obtained by astrometry.  The $\chi^2$ 
thus becomes a 7-parameter expression which is then minimised.  From now 
on, $i_C$ will refer to the inclination derived using that procedure.

\begin{figure}
\resizebox{\hsize}{!}{\includegraphics{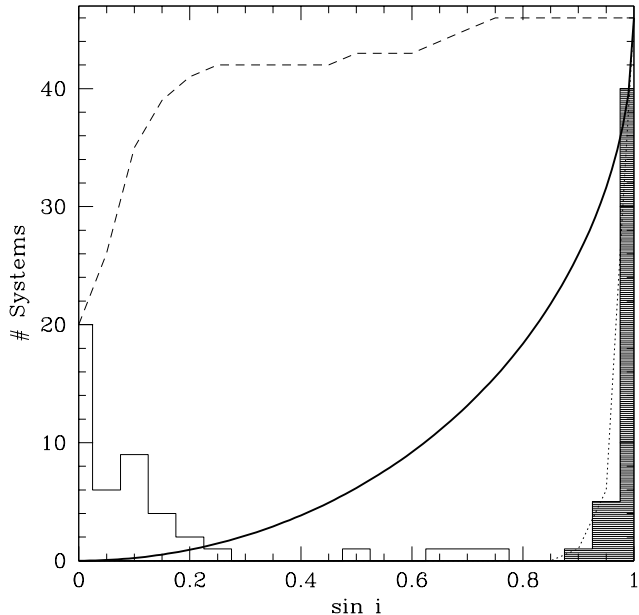}} 
\caption[]{\label{Fig:distri}Distribution of $\sin i_C$ ($\sin i_T$) 
plotted as open (filled) histograms.  The dashed and dotted lines 
represent the experimental cumulative frequencies of $i_C$ (Campbell) 
and $i_T$ (Thiele-Innes) respectively.  The thick continuous-line 
gives the theoretical cumulative frequency.}
\end{figure}

If the orbital planes of extrasolar companions are randomly oriented, 
Pr$(\sin i<x)=1-\sqrt{1-x^2}$.  However, when the Hipparcos data of 
46 systems\footnote{Based on the Catalog of extra-solar planets 
maintained by J.~Schneider (http://www.obspm.fr/encycl/catalog.html), 
content in early December 2000}  (Table \ref{Tab:systems}) are fitted 
using the above algorithm, the distribution of $i_C$ peaks close to 0 
(Fig.~\ref{Fig:distri}, open histogram).  In order to explain such a 
distribution, one can either argue that (a) the orbital planes of any 
unbiased sample are not randomly oriented; (b) there is a selection 
effect in the sample detected by radial velocity, pushing the inclinations
 towards small values, since the binaries have been excluded; (c) the 
inclinations thus derived are plain wrong.  We will no longer consider 
explanation (a): even if there could be a preferential inclination 
(e.g in clusters), that it could be 0 (i.e. towards our line of sight)
would rather be anthropocentric.

%
%

Figure \ref{Fig:distri} shows that the number of systems with small 
inclinations ($i_C$) is way too large with respect to the theoretical 
distribution.  For instance, there is only a 0.5\% probability
of getting a system with $\sin i<0.1$ whereas 32 systems over 
46 are characterised by $\sin i_C<0.1$.  

Could a selection effect of the radial velocity surveys explain this 
result, as suggested by \cite{Han-2001:a}?  To show that the answer is 
in the negative we may take account of all companions to solar-type 
primaries discovered through radial velocity studies and compute the 
probability of getting small inclinations.  Counting dwarf stars later 
than F5 from the spectroscopic binaries after \cite{Batten-1989:a} 
and \cite{Duquennoy-1991:b} and adding the recent low-mass 
companions gives about 450 stars, the exact number does not matter 
much as will be seen below.  On the average, we would expect 2 systems 
with $\sin i<0.1$ in our sample, so that the probability to get 32 
such objects is about $10^{-25}$. Even if we allowed for an underestimation
by a factor 10 (!) of the inclination, to get 16 systems with $\sin(10i)<0.1$
would have a $10^{-9}$ probability.

There may be unpublished spectroscopic companions, so let us consider 
{\it all} stars which can be observed.  From the Tycho Catalogue, 
there are 72\,162 stars with $V<9$ and $B-V$ between 0.4 and 1.4, 
corresponding to F5-K8, and only a small fraction of this population 
has been observed indeed.  Assuming a 11\% frequency of short period 
($P<1000$ days) spectroscopic binaries \citep{Halbwachs-2000:b}, one can 
expect on the average 0.4 system with $\sin i<0.01$ in this whole 
population whereas there are 16 systems in our sample (resp.  8 in the 
subsample of \cite{Han-2001:a}) i.e. a $10^{-6}$ (resp.  $10^{-2}$) 
probability.  Facing all these unrealistic probabilities, it is clear 
that the hypothesis (b) above is completely ruled out.

\cite{Arenou-2000:a} and \cite{Pourbaix-2001:a} favoured the explanation (c): 
the inclinations and their formal errors may be plain wrong.  The right-hand 
side of Eq.~\ref{Eq:defaasini} depends on the (known) spectroscopic elements 
and the parallax only: as shown by \cite{Halbwachs-2000:a}, $a_a$ is of the
order of the astrometric precision (about 0.5 mas for the considered stars).
$\sin i_C$ is then constrained by the spectroscopic $a_1 \sin i$ and the 
parallax: the smaller they are, the smaller the inclination.  So, the 
inclinations thus obtained could have no physical meaning and could 
just be artifacts of the fitting procedure.

%
\begin{figure}
\resizebox{\hsize}{!}{\includegraphics{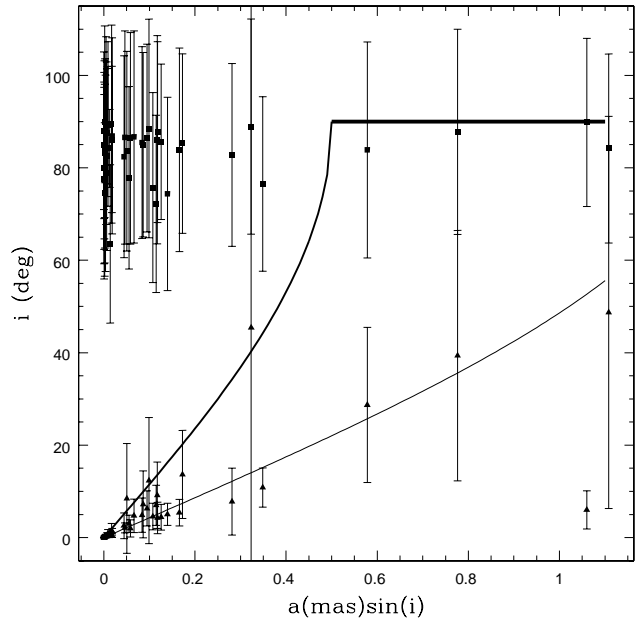}}
\caption[]{\label{Fig:ivsassini}Inclinations versus $a_a\sin i$ 
(Eq.~\ref{Eq:defaasini}). The triangles and squares denote $i_C$ 
and $i_T$ respectively.  The thick line gives the upper bound on $i_C$ 
based on $a_a\sin i$, assuming a semi-major axis of 0.5 mas. The thin line
is a least-square fit to the $i_C$ observed values.}
\end{figure}
%

%
\subsection{A test of the inclination}\label{Sect:incl}
%

In order to show that these $i_C$ are wrong, let us relax some 
constraints on the orbital parameters, namely $\omega_1$ and $K_1$.  
The orbital parameters $e$, $P$, and $T$ are set to the spectroscopic 
values whereas the others (advantageously replaced by the Thiele-Innes 
constants $A$, $B$, $F$, and $G$) are derived from the observations.  
One advantage of this procedure is that, once $e$, $P$, and $T$ are 
fixed, the least-square system of equations is linear with respect to 
the parameters and needs no iteration to converge.  Finally the 
parameters $\omega_1$, $a_a$, $i$ and $\Omega$ are computed from the 
Thiele-Innes constants.  We will denote by $i_T$ the inclinations thus 
obtained (squares in Fig.~\ref{Fig:ivsassini}).  If the observations 
do contain the astrometric signature of the planet, the two 
approaches, i.e. Campbell's and Thiele-Innes' ones, both 
mathematically correct, should yield consistent results.
We recall here that what we quote `Campbell's solution' is the 
determination of the Campbell parameters with the supplementary constraint
of $a_1\sin i$ and $\omega_1$ to their spectroscopic value. Without this
constraint the two quoted approaches are identical.

The uncertainties of $i_T$ were derived using Monte-Carlo simulation.  
Instead of comparing each pairs of inclinations, we compare the 
weighted means of the two sets (weight=$\sigma_i^{-2}$, and using $i$ 
in the first quadrant).  For the Thiele-Innes and Campbell sets we obtain 
respectively $\bar{i_T}=83^{\circ}\pm2.7^{\circ}$ and 
$\bar{i_C}=0.02^{\circ}\pm0.004^{\circ}$ ($6.6\pm 2$ for an 
unweighted mean).  The two means are clearly discrepant.

The distribution of $i_T$ is also given in Fig.~\ref{Fig:distri}.  
Although it looks closer to the theoretical distribution than $i_C$ 
does, one can nevertheless notice an excess of edge-on orbits and a 
deficit of intermediate inclinations.  The probability of getting no 
inclination below 60\degr\ in this sample is $10^{-10}$. These inclinations
are as unlikely as those obtained with the Campbell approach. 
Unfortunately, such a bias towards 90\degr\ could also be consistent 
with a selection effect due to the spectroscopic investigations.

Actually, random noise also yields edge-on Thiele-Innes orbits.  
Although, regardless of $e$, $P$ and $T$, fitting such a null signal 
leads to Thiele-Innes' elements normally distributed around 0 
(unbiased), $i_T(A, B, F, G)$, given by \citep{DoSt}
\begin{equation}
\tan^2(\frac{i_T}{2})=
\sqrt{\frac{(G-A)^2+(B+F)^2}{(A+G)^2+(B-F)^2}}\label{Eq:defi},
\end{equation}
follows a distribution centred in $\pi/2$.  When the standard 
deviations of the Thiele-Innes elements are all the same, 
$\sigma_i\approx11.7$\degr.

%
\begin{figure}
\resizebox{\hsize}{!}{\includegraphics{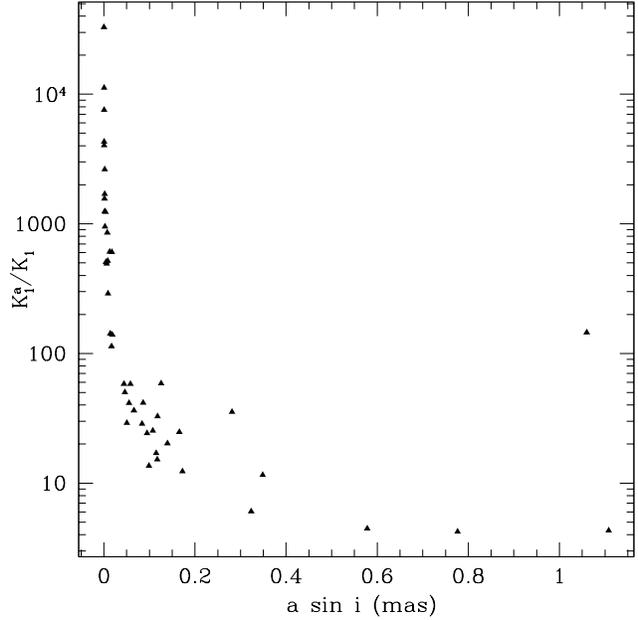}}
\caption[]{\label{Fig:ratioK}Ratio of the $K_1$ 
estimate based on the Thiele-Innes elements to the spectroscopic $K_1$.}
\end{figure}
%

However, even a likely $i_T$ would not mean the whole orbital solution 
is sensible.  Indeed, even if Campbell's elements are replaced with the 
Thiele-Innes ones, $a_a$ still remains closely related to the 
residuals of the coordinates of the star, the reason being that the 
computed semi-major axis roughly follows a law proportional to a 
$\chi(4)$ law, since $a_a\propto\sqrt{A^2+B^2+F^2+G^2}$ (from 
Eq.~\ref{eqM}).

If there was some orbital information in the Hipparcos data, one would 
expect a one to one relation between the astrometric $K_{1T}$ 
(computed using Eq.~\ref{Eq:defaasini}) and the spectroscopic $K_1$ 
(Fig.~\ref{Fig:ratioK}).  However the mean weighted difference between 
the two is $458\pm62$ m/s.  What happens here is that, with $a_a$ 
about 0.5 mas and an inclination $i_T$ close to 90\degr\, the 
astrometric estimate of $K_1$ is almost the inverse of the 
spectroscopic $a_1\sin i \times \varpi$.  Hence, what pushes the 
inclination towards zero with Campbell's approach when $a_1 \sin i$ 
gets small is also responsible for pushing the Thiele-Innes $K_1$ 
towards a large value.

%
\subsection{The law of the semi-major axis}\label{Sect:semi}
%

Since we cannot rely on the inclination, we now turn to the 
semi-major axis.  It is indeed their values together with a small
spectroscopic $a_1\sin i$ which pushed the inclinations towards small 
values in the Campbell approach.

\cite{Han-2001:a} selected the large secondary masses using the ratio 
$a_a\over{\sigma_{a}}$ so the question is whether large values of 
this ratio are expected or not.  We proceed by simulations 
since one cannot easily recover the true $a_1$ distribution either 
by assuming a Rayleigh-Rice law \citep{Halbwachs-2000:a} or with an 
approximation formula \citep{Han-2001:a}. The Rayleigh-Rice law was shown
by simulations to be globally compatible with the observed $a_1$ on a sample
of stars, but what we need here should be done on a star by star basis.

For each star, we assume the extreme case that all true $a_1$ are 
minimum, corresponding to $i=90$\degr, the other orbital and 
astrometric parameters being those obtained in the solution with the 
real data.  We then draw at random Gaussian Hipparcos abscissae around 
their expected value and a Campbell solution is computed.  Using 1000 such 
simulations we obtain the empirical probability for each star that 
the $a_a\over{\sigma_a}$ ratio is larger than the observed one 
(column labelled Pr$_a$ in Table \ref{Tab:systems}).

%
\begin{figure}
\resizebox{\hsize}{!}{\includegraphics{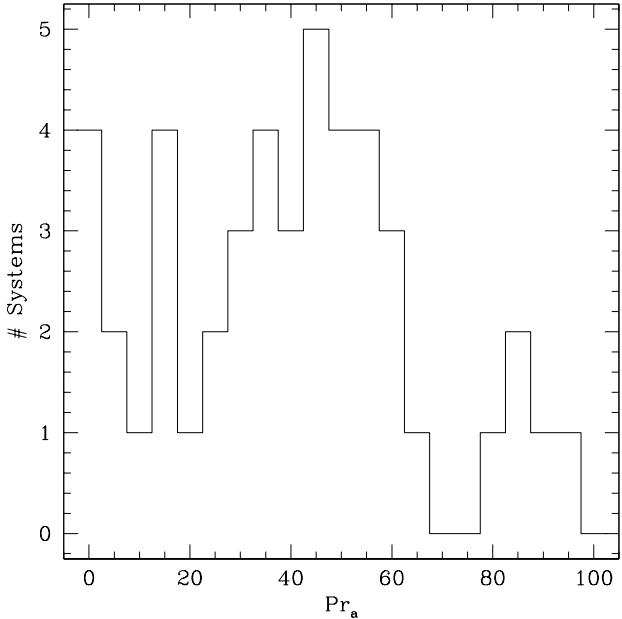}}
\caption[]{\label{Fig:unifa}Probability that simulated
$a_a\over{\sigma_a}$ (assuming $i=90$\degr) are larger than the 
observed one for the 46-sample.}
\end{figure}
%

As can be seen in Fig.~\ref{Fig:unifa}, this probability is not 
uniform (significance level 3\%) which is not surprising since the 
simulations assumed no outlier, no other unmodelled astrometric 
perturbation and $i=90$\degr.  The main point however is that there is 
no indication of a significant fraction of the real sample with a 
large ${a_a}\over{\sigma_a}$ ratio, which would correspond to small 
Pr$_a$ in the simulations.  There are six orbits with a probability 
smaller than 5\%, but this is not completely unexpected since this has 
a 3\% probability to occur in a 46-sample.  Among these stars, the 
result of \object{$\rho$ CrB} and \object{HD 195019} are however less 
likely (prob=0.01\%) to occur simultaneously.  This could indicate a 
significant orbit or other astrometric perturbations for these stars.

This exercise shows, in some sense, that a whole sample made of 
edge-on orbits would be more probable than the large fraction of 
face-on orbits which are obtained by the direct Campbell approach.  
Edge-on orbits for all stars are of course unrealistic, and this shows 
merely that there is no orbital information in the Hipparcos data for 
the sample as a whole.

%
\begin{table*}[htb]
\caption[]
{\label{Tab:systems}List of systems \citep{Pourbaix-2001:a} and the 
statistical tests.  
Pr$_a$=Probability to get a $a_a/\sigma_a$ larger than the observed 
value assuming $i$=90\degr; Pr$_1$=Pr($\hat{F_C}>F(2,N-7)$) 
\citep{Pourbaix-2001:a}; 
Pr$_2$=Pr($\hat{F}>F(4,N-9)$); Pr$_3$=Pr($\chi^2_{4S}>\chi^2(4)$); 
Pr$_4$=Pr($\chi^2_{4C}>\chi^2(4)$); The Campbell approach may be 
accepted when the conditions Pr$_1<5\%$, Pr$_2<5\%$, 
Pr$_3<5\%$, $|D|<\approx 2$, Pr$_4>5\%$, Pr$_5>5\%$ are fulfilled.}
\begin{tabular}{lll|r|rrrlrr}\hline
HIP & HD/name & Ref &Pr$_a$&Pr$_1$ & Pr$_2$ & Pr$_3$ & $D$ & Pr$_4$ & Pr$_5$\\
& & & (\%)& (\%) & (\%) & (\%) & & (\%) & (\%) \\ \hline
1292   & GJ 3021          & \cite{Naef-2000:a}      & 67 & 66 & 89 & 89 &  0.24 & 99 & 87\\
5054   & 6434             & \cite{Queloz-2000:a}    & 14 & 51 & 12 & 33 &  0.77 & 44 & 15\\
7513b  & $\upsilon$ And b & \cite{Butler-1999:a}    & 55 & 62 & 52 & 50 & -0.38 & 66 & 30\\
7513c  & $\upsilon$ And c & \cite{Butler-1999:a}    & 16 & 23 & 14 & 14 &  0.93 & 41 & 14\\
7513d  & $\upsilon$ And d & \cite{Butler-1999:a}    & 37 & 20 & 25 & 25 & -0.25 & 71 & 34\\
8159   & 10697            & \cite{Vogt-2000:a}      &  2 &  6 & 19 &  5 &  0.53 & 89 & 61\\
9683   & 12661            & \cite{Fischer-2001:a}   & 79 & 36 & 65 & 51 &  0.24 & 96 & 73\\
10138  & Gl 86            & \cite{Queloz-2000:b}    & 39 & 75 & 64 & 64 &  0.44 & 72 & 36\\
12048  & 16141            & \cite{Marcy-2000:a}     & 50 & 46 & 59 & 59 &  0.59 & 86 & 52\\
12653  & HR 810           & \cite{Kurster-2000:a}   & 28 & 20 & 28 & 23 &  0.34 & 95 & 73\\
14954  & 19994            & \cite{Queloz-2000:a}    & 42 & 40 & 60 & 61 &  0.21 & 92 & 63\\
16537  & $\epsilon$ Eri   & \cite{Hatzes-2000:a}    & 36 & 44 & 71 & 59 & -0.24 & 95 & 71\\
26381  & 37124            & \cite{Vogt-2000:a}      & 86 & 55 & 58 & 23 &  0.03 & 41 & 13\\
27253  & 38529            & \cite{Fischer-2001:a}   &  3 &  8 &  8 &  1 & -2.07 & 48 & 26\\
31246  & 46375            & \cite{Marcy-2000:a}     & 46 & 46 & 93 &  0 &  0.38 &  0 &  0\\
33719  & 52265            & \cite{Butler-2000:a}    & 54 & 59 & 77 & 74 &  0.11 & 93 & 66\\
43587  & 55 Cnc           & \cite{Butler-1997:a}    & 30 & 30 & 13 & 12 & -0.65 & 32 &  9\\
47007  & 82943            & \cite{Naef-2001:a}      & 25 & 61 & 68 & 61 &  0.04 & 81 & 46\\
47202b & 83443b           & \cite{Mayor-2000:a}     &  5 & 13 & 15 &  9 & -0.07 & 52 & 20\\
47202c & 83443c           & \cite{Mayor-2000:a}     &  2 &  3 &  8 &  4 & -1.04 & 77 & 40\\
50786  & 89744            & \cite{Korzennik-2000:a} & 86 & 35 & 62 & 42 &  0.60 & 93 & 69\\
52409  & 92788            & \cite{Queloz-2000:a}    & 26 & 22 & 31 & 27 & -0.17 & 74 & 38\\
53721  & 47 UMa           & \cite{Butler-1996:a}    & 97 & 86 & 98 & 96 & -0.12 & 99 & 90\\
64426  & 114762           & \cite{Marcy-1999:a}     & 56 & 41 & 52 & 57 &  0.21 & 86 & 52\\
65721  & 70 Vir           & \cite{Marcy-1996:a}     & 38 & 42 & 69 & 59 & -0.58 & 95 & 71\\
67275  & $\tau$ Boo       & \cite{Butler-1997:a}    & 60 & 55 & 74 & 34 & -0.72 & 25 & 14\\
68162  & 121504           & \cite{Queloz-2000:a}    & 21 & 27 & 47 & 35 & -0.51 & 87 & 54\\
72339  & 130322           & \cite{Udry-2000:b}      & 46 & 61 & 40 & 29 &  0.44 & 44 & 15\\
74500  & 134987           & \cite{Vogt-2000:a}      & 59 & 55 & 84 & 90 &  0.26 & 99 & 90\\
78459  & $\rho$ CrB       & \cite{Noyes-1999:a}     & .2 & .2 &  1 & .1 & -0.34 & 99 & 87\\
79248  & 14 Her           & \cite{Udry-2000:a}      & 36 & 25 & 37 & 44 &  0.19 & 86 & 45\\
89844b & 168443b          & \cite{Udry-2000:a}      & 44 & 50 & 96 & 77 &  0.22 & 95 & 75\\
89844c & 168443c          & \cite{Udry-2000:a}      & 88 & 60 & 92 & 63 & -0.17 & 98 & 91\\
90485  & 169830           & \cite{Naef-2001:a}      & 13 & 10 & 30 & 37 & -0.02 & 97 & 79\\
93746  & 177830           & \cite{Vogt-2000:a}      & 15 & 19 & 28 & 33 &  0.69 & 81 & 46\\
96901  & 16 Cyg B         & \cite{Cochran-1997:a}   & 44 & 41 & 65 & 67 & -0.52 & 95 & 70\\
97336  & 187123           & \cite{Vogt-2000:a}      & 34 & 17 & 46 & 62 & -0.22 & 99 & 95\\
98714  & 190228           & \cite{Sivan-2000:a}     & 12 &  9 & 24 & 22 & -0.27 & 94 & 67\\
99711  & 192263           & \cite{Vogt-2000:a}      & 30 & 73 & 48 & 20 & -2.48 & 30 &  8\\
100970 & 195019           & \cite{Vogt-2000:a}      & .5 &  0 &  0 &  0 & -4.68 &  6 &  1\\
108859 & 209458           & \cite{Mazeh-2000:a}     & 48 & 19 & 15 &  8 & -2.08 & 40 & 13\\
109378 & 210277           & \cite{Vogt-2000:a}      & 53 & 41 & 72 & 66 &  0.00 & 99 & 97\\
113020 & Gl 876           & \cite{Delfosse-1998:a}  & 49 & 49 & 43 & 53 &  0.32 & 74 & 37\\
113357 & 51 Peg           & \cite{Mayor-1995:a}     & 46 & 65 & 86 & 79 & -0.13 & 99 & 95\\
113421 & 217107           & \cite{Vogt-2000:a}      & 61 & 54 & 60 & 68 &  0.62 & 86 & 53\\
116906 & 222582           & \cite{Vogt-2000:a}      & 52 & 10 & 17 & 28 &  0.32 & 82 & 47\\
\hline
\end{tabular}
\end{table*}
%

%
\section{Screening the astrometric solutions}\label{Sect:screen}
%

At this point, it is clear that there is no indication in the 
planetary sample as a whole of a significant fraction of low 
inclinations/large secondary mass.  Due to the increasing number of 
planetary candidates, it is however expected on a statistical basis 
that a bunch of them may have a mass much larger than the minimum 
mass.

Since there is evidence that a direct use of the orbital parameters 
(without taking their statistical distribution into account) may lead 
to incorrect conclusions, we need a set of tests which can tell us
whether the Campbell results  are correct or not. Since by construction 
the Campbell analysis is consistent with the spectroscopic orbit, 
this is where the Thiele-Innes approach reveals useful.

The tests described below intend to show whether an orbital model 
improves the fit of the Hipparcos data or not, whether the 
Thiele-Innes elements are consistent with Campbell's ones or not and, 
finally, whether the astrometric and spectroscopic orbits agree with 
each other or not.  This is an essential point: before combining the 
spectroscopic and astrometric data, as is done in the Campbell 
approach, one has to be sure that both solutions are compatible.  If 
all these tests give a positive result, then the orbital elements 
obtained with the Campbell approach may be trusted.

In the other case, one may either conclude that the orbital elements 
were obtained by chance only, or that there is a perturbation in the 
astrometric solution due e.g. to another companion.  The presence of 
other companions around several stars hosting planetary candidates can 
be detected from a radial velocity residual trend 
\citep{Fischer-2001:a} and some of these companions are being 
discovered by adaptive optics: \object{HD 114762}, \object{$\tau$ 
Boo} \citep{LLoyd-2000:a} and \object{Gl 86} \citep{Els-2001:a}.  
\cite{Arenou-2000:a} combined the spectroscopic data with the 
Hipparcos and Tycho2 data \citep{Hog-2000:a} for the planetary 
candidates and the use of long term proper motion gave a hint of 
possible other companions (period larger than several decades).

The spectroscopic orbital elements are generally precise.  So, from now
on, we treat them as constants.  As we 
have seen, the statistical distributions of the semi-major axis and of 
the inclination are not trivial, so our tests must be based on better 
known parameters.  This is the case for the $A, B, F, G$ Thiele-Innes 
parameters: obtained by a linear least-squares solution, they may be 
considered Gaussian around 0, as has been confirmed by direct 
simulations.  Concerning real Hipparcos data, it has been shown that 
the errors on the astrometric parameters could be considered Gaussian 
\citep{Arenou-1995}.  So the normality hypothesis for the Thiele-Innes 
elements is not just a convenient hypothesis but is plainly justified.
Of course, a departure from normality may be due to outliers.  Apart 
from possible instrumental errors, however, outliers may be normal 
observations when a good orbital model is used in place of the 
standard ``single star'' model adopted for the major part of the 
Hipparcos stars.

%
\subsection{Need for an orbital motion}
%

As pointed out by \cite{Pourbaix-2001:a}, none of the Campbell-like 
orbital solutions but four does improve the fit of the Hipparcos 
intermediate astrometric data (IAD).  Could it be different with a 
Thiele-Innes solution?  Do the two additional degrees of freedom 
significantly reduce the $\chi^2$?  The quantity
\begin{equation}
\hat{F}=\frac{N-9}{4}\frac{\chi^2_S-\chi^2_T}{\chi^2_T}\label{Eq:Fhat}
\end{equation}
follows a F-distribution with $(4,N-9)$ degrees of freedom 
\citep{DaReErAnPhSc}.  $N$ denotes the number of data points and 
$\chi^2_S$ and $\chi^2_T$ are the value of the $\chi^2$ with the 
5-parameter (single star) model and Thiele-Innes' orbital model 
respectively.  We reject the null hypothesis `no orbital wobble 
present in the IAD' if the probability of getting $\hat{F}$ larger 
than $F(4,N-9)$ is lower than 5\%.  This P-value is given   
as Pr$_2$ in Table \ref{Tab:systems} where we also recall as 
Pr$_1$ the analogous probability obtained by \cite{Pourbaix-2001:a}
for the Campbell approach.

%
\subsection{Significance of $A$, $B$, $F$, and $G$}
%

Getting a substantial reduction of the $\chi^2$ with the Thiele-Innes 
model does not necessary mean the four constants are significantly 
different from 0.  In order to assess that, we compute
\begin{equation}
\chi^2_{4S}=\mathbf{X}^{\rm t}\mathbf{V}^{-1}\mathbf{X}\label{Eq:chi24S}
\end{equation}
where $\mathbf{X}=(A, B, F, G)$ and $\mathbf{V}$ is the covariance 
matrix of $A$, $B$, $F$, and $G$.`${\rm t}$' denotes the 
transposition.  $\chi^2_{4S}$ follows a $\chi^2$-distribution with 4 
degrees of freedom when there is no orbital information.  When all 
formal errors on Thiele-Innes parameters are identical, this 
$\chi^2_{4S}$ is proportional to $a_a$.  We then 
reject the null hypothesis `$a_a$ (mas) is significantly different 
from 0' if the probability that $\chi^2_{4S}$ exceeds $\chi^2(4)$ is 
less than 5\%.  This P-value Pr$_3$ is given as in column 7 of 
Table \ref{Tab:systems}.

%
\subsection{Consistency between spectroscopy and astrometry}
%

Testing such a consistency is a bit tricky.  Indeed, the nature of the 
information supplied with by both sides is essentially different.  The 
astrometry should however be able to recover $a\sin i$ and $\omega_1$, 
both given by spectroscopy.  There are a few useful relations based on 
the Thiele-Innes elements such as

\begin{eqnarray}
\qquad L&=&A^2+B^2-(F^2+G^2)=(a_a\sin i)^2\cos(2\omega_1)\label{eqL}\\
\mbox{and}\,\,
M&=&A^2+B^2+F^2+G^2=2a_a^2-(a_a\sin i)^2\label{eqM}
\end{eqnarray}
that can be used to build up a test.  For instance,
\begin{equation}
D=\frac{L_l-L_r}{2\sigma\sqrt{M_r+2\sigma^2}}\label{Eq:defD}
\end{equation}
follows a distribution of null expectation and unit variance.  $L_l$ 
denotes the left-hand side of the definition of $L$, i.e. based on the 
Thiele-Innes constants only while $L_r$ and $M_r$ are computed from 
the spectroscopic elements and the parallax.  In the latter, the true 
$a_a$ is unfortunately unknown, so it is derived from the 
Campbell-like solution.  $\sigma$ stands for the mean standard errors 
of $A$, $B$, $F$, and $G$.  Because we neglect the correlation between 
the Thiele-Innes constants, and with the $a_a$ deduced from the 
Campbell analysis, $D$ is not Gaussian, and $|D|<2$ does not strictly 
correspond to a 95\% confidence level.  Indeed, as can be seen in 
Table \ref{Tab:systems}, the distribution of $D$ is leptokurtic: $a_a$ 
being overestimated, $|D|$ is generally underestimated.  A large value 
of $|D|$ is then a good indication that the astrometric solution is 
not consistent with the spectroscopic orbit.

%
\subsection{Consistency between Thiele-Innes' and Campbell's solutions}
%

Since both methods are applied to the same data set, they should also 
give similar results.  We have already shown that edge-on orbits 
are consistent with pure noise signal when fitted with the 
Thiele-Innes parameters while \cite{Pourbaix-2001:a} got similar 
results for face-on orbits with Campbell's elements.

Here also, designing a test may be difficult, especially because the 
distribution of $\Omega$ is anything but symmetrical.  So, instead of 
comparing the two sets of Campbell elements, we compute their 
Thiele-Innes expressions.  We then build
\begin{equation}
\chi^2_{4C}=\mathbf{Y}^{\rm t}\mathbf{V}^{-1}\mathbf{Y}\label{Eq:chi24C}
\end{equation}
where $\mathbf{Y}=(A-A_C, B-B_C, F-F_C, G-G_C)$, $\mathbf{V}$ has the same 
meaning as in $\chi^2_{4S}$ (Eq.~\ref{Eq:chi24S}) and $(A_C, B_C, F_C, 
G_C)$ are computed from the Campbell elements.  If we neglect the 
correlations between Campbell and Thiele-Innes, the quantity 
$\chi^2_{4C}$ follows a $\chi^2$-distribution with 4 degrees of 
freedom.  We reject the null hypothesis `Thiele-Innes' and Campbell's 
approaches yield consistent solutions' if the probability that 
$\chi^2_{4C}$ exceeds $\chi^2(4)$ is less than 5\%.  That probability 
Pr$_4$ is given in Table \ref{Tab:systems}.  Because of the 
correlations between the $A, B, F, G$ obtained by both methods, a 
rejection at a 5\% level will occur in much less than  
5\% of the cases.  In fact, $\chi^2_{4C}$ is probably closer to a 
$\chi^2$-distribution with 2 degrees of freedom, so this test is less 
discriminant than the next one.

%
\subsection{A small inclination}
%

Finally, one may answer both whether a small inclination is present or 
whether the astrometric solution is compatible with the spectroscopic 
one, still using the Thiele-Innes elements as an external control. 
Specifically,
\begin{eqnarray}
A-G&=&\varpi a_1\sin i ~ \cos(\Omega-\omega_1) ~ \tan ({i\over 2})\nonumber\\
B+F&=&\varpi a_1\sin i ~ \sin(\Omega-\omega_1) ~ \tan ({i\over 2})\label{AmB}
\end{eqnarray}
and
\begin{eqnarray}
A+G&=&\varpi a_1\sin i ~ \cos(\Omega+\omega_1) ~ \cot ({i\over 2})\nonumber\\
B-F&=&\varpi a_1\sin i ~ \sin(\Omega+\omega_1) ~ \cot ({i\over 2})\label{ApB}
\end{eqnarray}

In the case of the planetary candidates, $a_a\sin i$ is small, and, if 
the inclination is really small, then the right-hand sides of 
Eqs.~\ref{AmB} are null; when the inclination is found to be near 
180\degr, the Eqs.~\ref{ApB} should be $\approx 0$.  We test whether 
Eqs.~\ref{AmB} (if $\cos i_C>0$) or Eqs.~\ref{ApB} (if $\cos i_C<0$) are 
verified.  If $\mathbf{Z}$ denotes this two-dimensional vector, and 
$\mathbf{W}$ its covariance matrix computed using the covariance 
matrix of $(A,B,F,G)$, then
\begin{equation}
\chi^2_{2}=\mathbf{Z}^{\rm t}\mathbf{W}^{-1}\mathbf{Z}\label{Eq:chi22}
\end{equation}
should follow a $\chi^2$ with two degrees of freedom under the null 
hypothesis of a Thiele-Innes solution compatible with the 
Campbell/spectroscopic solution.  The P-value Pr$_5$ is given in 
Table~\ref{Tab:systems}. We will reject the compatibility between
Thiele-Innes and Campbell if Pr$_5$ is smaller than 5\%.

%
\subsection{Discussion}
%

Since the proposed tests are not independent, we cannot compute an 
overall probability.  While the first three tests check whether the 
astrometric solution is firmly established, the last three ones 
(useless in the case of a `noise only' orbit) assess the agreement 
between the astrometric and spectroscopic orbits.  In fact Pr$_1$ 
and Pr$_5$ would probably be enough for investigations of future 
planetary candidates, provided that a signal is indeed present at the
spectroscopic period.  It appears that, with the exception of $\rho$ 
CrB, almost all stars in the present sample fail to at least two of 
the proposed tests.

Under the assumption of a `noise only' orbit, one expects that the 
computed probabilities are uniform on this sample.  Once again, a 
departure of the Hipparcos residuals from strict normality (e.g. due 
to outliers or long period companions) could produce spurious 
solutions which in turn would contaminate the probabilities.  
Nevertheless, using a Kolmogorov test, the uniform null hypothesis is 
accepted for Pr$_2$ (P-value=88\%), Pr$_3$ (7\%), Pr$_5$ (98\%).

If this was necessary, this terminates the demonstration that the 
Hipparcos data mainly contains noise.  This does not mean that there 
is no brown or red stars unfortunately interpreted as planet (apart 
$\rho$ CrB), but rather that Hipparcos is of little help to demonstrate 
it, to the contrary of the \cite{Han-2001:a} analysis.

%
\section{Individual stars}
%
%
\subsection{Validity of $i_C$ with HD~209458}\label{Sect:HD209458}
%

HD~209458 (HIP~108859) is, so far, the only case where the inclination 
is accurately known.  Indeed, transits of the planet have been 
monitored since 1999 \citep{Charbonneau-2000:a}, thus setting a strong 
constraint on the inclination: $i=86.1\pm 1.6$\degr\ \citep{Mazeh-2000:a}.

Hipparcos observed that system and it has been noticed later on 
that it did detect the planet \dots\ thanks to its photometric 
signature \citep{Soderhjelm-1999:a,Robichon-2000:a, 
Castellano-2000:a}.  However, that result does not mean the 
astrometric signature is also in the IAD. Thus, when $i_C$ and 
$\Omega$ are fitted together with the five astrometric parameters, the 
inclination is $i_C=0.019\pm0.0097$\degr.  Fitting the Thiele-Innes 
elements yields an inclination $i_T=76\pm18$\degr.

It appears that the `noise only' null hypothesis can't be rejected as 
indicated by three tests (Pr$_1$, Pr$_2$ and Pr$_3$).  Moreover the 
hypothesis of a small inclination has a small P-value (Pr$_5$=13\%).  
On this real case where the true orbit is known, the proposed tests 
thus seems discriminating enough to detect wrong Campbell solutions.

%
\subsection{$\rho$ CrB}\label{Sect:rhoCrB}
%

The suggestion that the secondary could in fact be an M dwarf was made 
by \cite{Gatewood-2001:a} using both Hipparcos and MAP \citep{Gatewood-1987:a}
data, and the solution in the Hipparcos Catalogue was already detected as 
orbital with a period twice the spectroscopic period.  As shown by 
\cite{Gatewood-2001:a} a periodogram exhibits a strong signal at this 
period.

This is the only star in the `planetary' sample which succeeds to all 
tests: while the hypothesis of a `noise only' orbit is significantly 
rejected by the three first tests, there is no indication that the 
spectroscopic, Campbell and Thiele-Innes orbits could be inconsistent, 
using the other tests.  Consequently, the main result of 
\cite{Gatewood-2001:a} may not be rejected.

%
\subsection{HD~10697}\label{Sect:HD10697}
%

Since the minimum reflex semi-major axis is about $0.4$ mas, and thus 
suggests that Hipparcos could give a hint of the true semi-major axis, 
\cite{Zucker-2000:a} combined the astrometric and spectroscopic data 
and obtained $a_a=2.1\pm0.7$ mas, corresponding to a $38\pm 13$ M$_J$ 
mass for the secondary. This result would indicate that the secondary is in
the brown dwarf domain.

%
\begin{table}[b]
\caption[]
{\label{Tab:trial}Campbell analysis with different trial periods
on HD~10697}~\label{Tab:Per10697}
\begin{tabular}{r|r|r} \hline
Period  & $a_a$ &GOF\\
(days)&(mas)&\\ \hline
 700&$1.9\pm 0.7$&1.31\\
 800&$1.9\pm 0.7$&1.28\\
 900&$1.9\pm 0.7$&1.27\\
1000&$2.\pm 0.7$&1.32\\
1072&$2.1\pm 0.7$&1.28\\
1100&$2.2\pm 0.8$&1.25\\
1200&$2.7\pm 0.9$&1.24\\
1300&$3\pm 1$&1.29\\
\hline
\end{tabular}
\end{table}
%

\begin{figure}
\resizebox{\hsize}{!}{\includegraphics{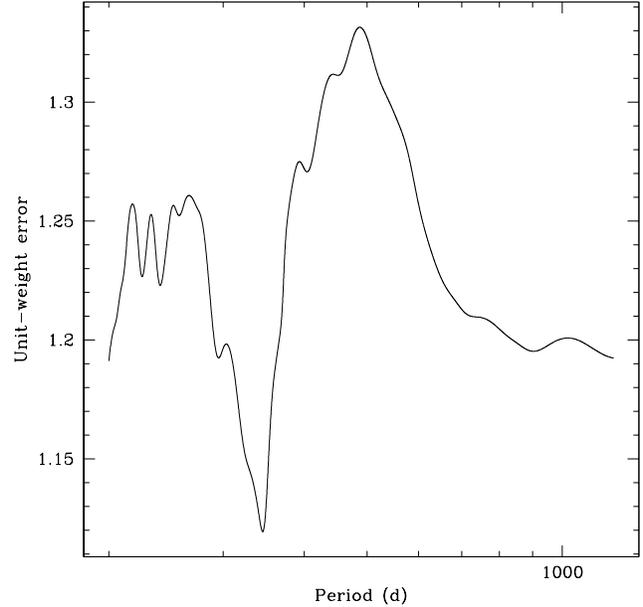}}
\caption[]{\label{Fig:Per10697base}Periodogram of HD~10697 assuming $e$ and
$T$ from the known companion.}
\end{figure}

Although the signal should be present to some extent in the Hipparcos 
data, a periodogram based on the unit-weight error of the Thiele-Innes
solution shows no signal at the 1072 days spectroscopic period
(Fig.~\ref{Fig:Per10697base}). Instead a 345 days period would be more
likely, but would have been seen in the spectroscopic data.

Repeating this periodogram with the Campbell analysis is shown in 
Table~\ref{Tab:Per10697}.  While the obtained semi-major axis increases with 
trial period, the inclination decreases accordingly, the same large signal to
noise $\approx 3$ is obtained, and the goodness of fit (GOF) is not improved 
at the spectroscopic period. The Campbell analysis simply adjusts itself
to the constrained spectroscopic elements, without showing any 
sensible information in the astrometric data.

This explains the negative results of our tests in 
Table~\ref{Tab:systems}.  Due to the size of the astrometric residuals 
(partly independent from the orbital perturbation), the obtained 
semi-major axis is probably overestimated, but it is difficult to know 
to which extent. Consequently the exact mass of the secondary 
cannot be obtained at face value using a combination of the Hipparcos 
data with the spectroscopic orbital elements.

%
%

%
\subsection{HD~195019}\label{Sect:HD195019}
%

\begin{figure}
\resizebox{\hsize}{!}{\includegraphics{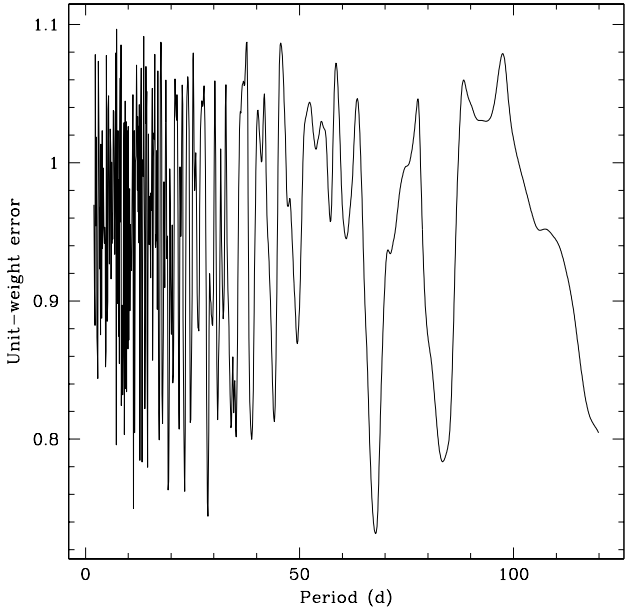}}
\caption[]{\label{Fig:Per195019base}Periodogram of HD~195019 assuming $e$ and
$T$ from the known companion.}
\end{figure}

This is probably a typical case of how one can be mislead by a direct 
Campbell analysis: the fit of the IAD is significantly improved by the 
orbital model but fails the last two tests, i.e. the astrometric orbit 
is clearly not consistent with the spectroscopic one.  The periodogram 
based on the orbital parameters of the known companion 
(Fig.~\ref{Fig:Per195019base}) reveals a peak at 68 days and nothing 
around 18 days (the orbital period of the planetary companion).

We cannot find any satisfactory solution by fitting the 12 parameters with
a global optimisation technique.  We end up with a minimum $\chi^2$ which
corresponds to $e\approx 1$ and a total correlation between $F$, $G$, and $e$.
Even if that solution does minimise $\chi^2(12)$, it is physically very 
unlikely.

%
%

Instead of fitting the 12 parameters in once, one can use the 
periodogram and build up the solution by adding the parameters one by 
one.  Thus, by first fixing $P$ to 67.935d, one derives 
$T_0=JD~2,448,296.8$ which yields the lowest value of the unit-weight 
error.  The resulting orbit would have an inclination of 
$65\pm18$\degr\  and $K_1=24\pm8$km/s.  Unlike the orbit of the known
companion, this solution successfully passes the last two tests.  However 
such a solution would have been present in the radial velocity residuals.

Another explanation could originate in the fact that HD 195019 is 
known since 1881 as a binary star, with a B component 4.5" apart 
(3.5" in 1988) with a position angle of 323\degr (resp.  330\degr), 
from the WDS Catalog.  Despite this fact, it has been dealt as a single 
star in the Hipparcos Catalogue (although a `suspected non-single
star' flag is set), because no convincing double star solution could be 
used.  Although $\approx 3$ mag fainter, the influence of the 
secondary may have produced astrometric residuals (the projected 
separation varying with the satellite scan orientation) which could be 
wrongly attributed to the reflex motion due to the low-mass companion,
especially if the companion had a significant motion during the 
Hipparcos observation period.

%
\subsection{HD~38529}\label{Sect:HD38529}
%

A periodogram reveals nothing special at the 14-day spectroscopic 
period.  However, this is another star for which \cite{Han-2001:a} 
suggest that the secondary could be an M dwarf.  They also quote that 
this star received an acceleration solution in the Hipparcos 
Catalogue. 

Before looking to the small signal due to the spectroscopic companion, 
one has to remove the effect due to the long period companion.  When 
the acceleration terms in time are thus subtracted, the semi-major 
axis of the Campbell solution becomes $1.3\pm .7$ mas, i.e. not 
significant, whereas it was $2.1\pm .6$ mas before; when a cubic 
instead of quadratic term in time is accounted for, $a_a=.9 \pm .7$ 
mas.

The influence of a long-period companion, also detected in the 
spectroscopic data \citep{Fischer-2001:a}, is thus a much more natural 
explanation to the size of the Hipparcos residuals than a tiny 
inclination for the short period companion.

%
\subsection{HD~83443}\label{Sect:HD83443}
%

\begin{figure}
\resizebox{\hsize}{!}{\includegraphics{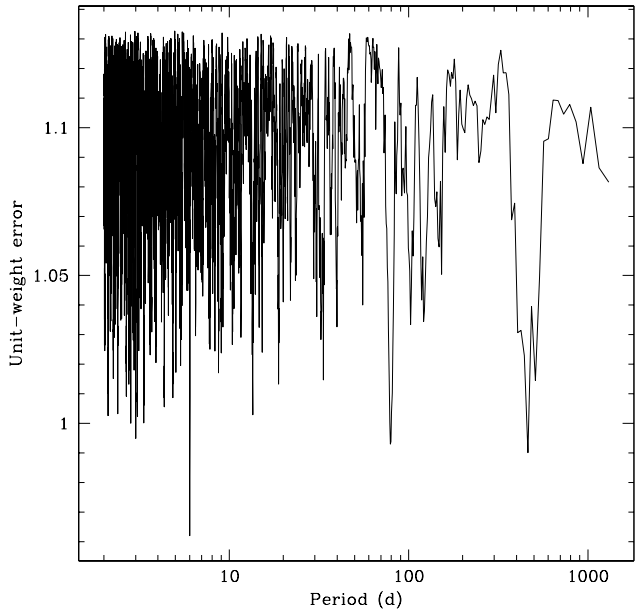}}
\caption[]{\label{Fig:Per47202}Periodogram of HD~83443 assuming $e$ and
$T$ from the second companion.}
\end{figure}

There is also no signal around the periods of the two planetary 
candidates ($\approx 3$ and 30 days) in the Hipparcos data.  Instead a 
period of 6, 79 or 462 days (Fig.~\ref{Fig:Per47202}) would be much more 
likely, but it would clearly have been seen in the spectroscopic data.
An unmodelled astrometric perturbation due to a longer period companion
would provide a more logical explanation, since there is a possible trend 
in the radial velocity residuals (S.~Udry, private communication).

%
%

Still, based on the tests we presented, the astrometric solution for 
HD~83443c was about to be accepted. The reason for this is the large
$a_a=2\pm 0.7$ mas. 

One may ask whether {\it any} other period could have given the same 
signal as the one obtained with a 30 days period.  All above 
mentioned tests were based on normality assumptions about the 
astrometric residuals but another, non-parametrical test may also be 
used.  A periodogram, based on the amplitude of the signal 
(root-sum-square of the 4 Thiele-Innes parameters) has been done: 
using 5000 trial periods between 2 and 1500 days, equally spaced in 
frequency, a Thiele-Innes solution is performed.  It happens that 30\% 
of the trial periods give an amplitude larger than the one obtained at 
the period of the spectroscopic companion.  In other words, the 
astrometric result has no special meaning.

%
\section{Conclusion}
%

In summary, the available Hipparcos data is not precise enough to
show that a significant part of the planetary candidates 
could be of stellar nature.

Although we expect a very small number of small inclinations in the 
putative extrasolar planets sample, the inclinations derived using the 
Campbell elements together with the Hipparcos data are pushed towards 
unreasonably small values.  As we have shown, this is not due to an 
observational selection but rather to the employed reduction method and
probably also due to the presence of longer period, unresolved companions. 

Though the Thiele-Innes alternative method gives inclinations more 
in line with what is generally expected from the spectroscopic 
detection, the estimated astrometric semi-major axis still does not 
provide a statistically unbiased estimate of the true reflex motion.  
Indeed, in the case of pure noise in the astrometric data, the 
Campbell approach would give inclinations biased towards 0\degr\ while 
Thiele-Innes recovers inclinations biased towards 90\degr.  So, none 
of the two methods can be trusted, except when it can be shown that 
they agree together, which is the basis of one of the tests we 
propose.  The other tests determine whether the astrometric orbit is 
significant or might result from noise only.

\newcommand\ignore[1]{} It must be pointed out that the bias on the 
semi-major axis and inclination only occurs for negligible orbital 
wobbles.  When a true orbital information is present in the Hipparcos 
astrometric data, then the orbital parameters are much better behaved.  
For instance, we may question whether the rejection of putative brown 
dwarfs into the stellar domain by the \cite{Halbwachs-2000:a} study 
was justified.  The stars \object{HIP 13769}, \object{\ignore{HIP 
}19832}, \object{\ignore{HIP }62145}, \object{\ignore{HIP }63366} and 
\object{\ignore{HIP }113718} which were found with a mass above the 
H-burning limit with a $2\sigma$ significance and \object{HIP 77152}, 
one $\sigma$ above, have been tested using the tests described above.  
All these stars succeed to all tests, apart from \object{HIP 19832} 
which fails to the last test (significance=3\%, this may be by chance 
only since the reality of the orbit is clear, as seen e.g. 
by periodogram).  The main conclusion of 
\cite{Halbwachs-2000:a}, the deficit of short period brown dwarf 
secondaries, is thus confirmed by this analysis.

In contrast, in the sample of putative planets, there is no indication 
that a significant fraction of secondaries could be brown or red 
dwarfs, with the possible exception of $\rho$ CrB. Is that 
to say that the Hipparcos astrometry is useless for the extrasolar 
planet study?  Not completely, since at least an upper limit on the 
secondary mass may be obtained, and Hipparcos may also give a hint of 
other, longer period companions.  For a satellite which was originally 
planned to get positions, proper motions and parallaxes, this is not 
negligible.

\begin{acknowledgements}
We thank J-L. Halbwachs, M. Mayor and S. Udry for their useful comments.  
DP thanks the National Aeronautics and Space Administration which 
partially supported this work via grant NAG5-6734.
\end{acknowledgements}

%
\bibliographystyle{apj}
\bibliography{articles,books}
%

\end{document}